# Advanced Lanczos diagonalization for models of quantum disordered systems

I. Kh. Zharekeshev and B. Kramer

I. Institut für Theoretische Physik, Universität Hamburg,
Jungiusstr. 9, 20355 Hamburg, Germany

An application of an effective numerical algorithm for solving eigenvalue problems which arise in modelling electronic properties of quantum disordered systems is considered. We study the electron states at the localization-delocalization transition induced by a random potential in the framework of the Anderson lattice model. The computation of the interior of the spectrum and corresponding wavefunctions for very sparse, hermitian matrices of sizes exceeding $10^6 \times 10^6$ is performed by the Lanczos-type method especially modified for investigating statistical properties of energy levels and eigenfunction amplitudes.

*Introduction.* The Lanczos algorithm is one of the most effective computational tools for searching few extreme eigenvalues and corresponding eigenvectors of large sparse hermitian matrices. It is successfully applied to many problems in atomic, molecular and condensed matter physics, where it is required to gain the information about low-lying excitations in a spectrum close to the ground state. In many applications the matrices are sparse due to various constraints, for example, a limited connectivity and a short-range coupling. Practically, the number of non-zero matrix elements is often proportional to the dimension of the matrix $N$ rather than to $N^2$.

Due to the quantum interference effects the one-electron states in the presence of a random impurity potential can be localized in a finite range of space. The localization phenomenon is an origin of the disorder-induced metal-insulator transition. This phase transition is of particular interest from the viewpoint of studying the spatial structure of the wave functions and the statistical properties of the spectrum at the critical point. It was recently found [1] that the correlations between the discrete energy levels at the mobility edge possess a *universal* property in the sense that they do not depend on the size of the system. In other words, the statistics of eigenenergies calculated for finite systems can be considered as a good approximation of those in the thermodynamic limit. At the same time the electron states at the transition are neither localized nor delocalized, but they become multifractal [2]. That is why it is important to study critical behavior of eigensolutions by using direct large-scale numerical diagonalization.

*Model and optimization schemes.* One of the common models which is used for describing electron transport properties in disordered systems close to the transition is the Anderson model of localization. The Hamiltonian of the model defined on a three-dimensional lattice is given by the operator $\hat{H} = \sum_n \epsilon_n c_n^\dagger c_n + t \sum_{n \neq m} (c_n^\dagger c_m + c_n c_m^\dagger)$, where $c_n^\dagger$ ($c_n$) is the creation (annihilation) operator of an electron at a lattice site $n$, and $m$ labels the sites neighboring to the site $n$. The on-site energies $\epsilon_n$ are random variables which are uniformly distributed within an interval of width $W$, such that $W$ parameterizes the disorder. The second term corresponds to the hopping processes between the neighbour sites in the lattice. The transfer integral $t$ can be set to unity ($t = 1$). The metal-insulator transition in the band centre, $E=0$, occurs at the disorder $W = W_c \approx 16.4$.

The sparse structure of the matrix $A$ corresponding to the Hamiltonian $\hat{H}$ makes the Lanczos method the most suitable for solving the eigenvalue problem $\hat{H}\psi_n = E_n \psi_n$ of large systems. For instance, the number of non-zero off-diagonal elements per line equals six in a simple cubic lattice of linear size $L$. Indeed, this speeds up substantially the most time-consuming part of the implementation, namely the matrix-vector multiplications. A single matrix-vector product needs computer time proportional to $N = L^3$. In addition, by applying the helical boundary conditions instead of traditionally used periodic boundary conditions, as was made in [2], it is possible to eliminate several expensive loops of the program. Another advanced optimization is to reorganize the position of the co-diagonals, so that some of them go beyond the borders of the original square matrix $A$, becoming as a continuation of other co-diagonals. This leads to a rectangular matrix of size $(L^2+N) \times N$. All of the off-diagonal matrix entries are presented in the form of six co-diagonals of the same length $N$ parallel to the main diagonal and horizontally shifted from it by the distances $-L^2, -L, -1, 1, L, L^2$, respectively. This restructuring of the matrix $A$ does not alter components of the resulting matrix-vector product. In fact, it allows to avoid the procedure of 'look-up' of the table-given positions of the $A$-matrix elements, which are usually necessary in the standard Lanczos diagonalization of the Anderson Hamiltonian [1, 3]. As a consequence, the memory requirements are also markedly diminished. Thus, instead of storing the initial structure of the matrix in the main processor memory, one needs only the fast random-number generator in order to create repeatedly the same set of the diagonal elements for each matrix-vector multiplication. (Details of the advanced steps will be published elsewhere.)

The convergence process to the eigensolutions was controlled by determining the upper limit of the user-specified tolerance $\delta = ||H\psi_n - E_n\psi_n||$. It is known [4] that due to the floating point arithmetic, which leads to the round-off errors, the Lanczos vectors lose their global orthogonality. Therefore, after the recursions finish the size $M$ of the tridiagonal matrix ($T_M$-matrix [4]) approaches $M \approx (1.9 - 2.2)N$, as shown in Fig. 1. This size is required in order to detect the pairs of the eigenvalues and the corresponding eigenvectors with the absolute precision $\delta \leq 10^{-9}$ in the internal part of the spectrum, where the density of states is maximal. Closer to the energy band edge the optimal value of $M$ drastically decreases. Stability of the modified Lanczos-type algorithm has been tested in a wide range of the system size $L$ ranging from 5 to 100.

There is a specific property of the spectrum of disordered systems at criticality, which turns to be important and helpful for the converging procedure. It is a strong quantum-mechanical level repulsion that decreases the probability to find two eigenvalues close to each other with decreasing the spacing between them. This property facilitates not only the speed of the iterations, but also the bisection algorithm for finding the eigenvalues of the $T_M$-matrix. The latter becomes more time-consuming, when one needs to calculate relatively large part of the spectrum (more than 10% of all of the eigenvalues). In this case the performance has further been improved by parallelizing the bisection algorithm for consequent energy sub-intervals. By varying the disorder $W$ apart from the critical value, we have checked that the size of the $T_M$-matrix should be slightly larger in order to reach the desirable accuracy of eigenvalues in a given part of the spectrum.

*Results.* By using the above optimizations it was possible to compute not only the extreme eigenvalues in the tails of the density of states, but also the interior of the spectrum of the disordered lattices of large size $N=100\times100\times100$ [5]. The maximal dimensions of the matrices from previous diagonalization studies [1, 3] have been exceeded by more than two orders of magnitude. The spectral fluctuations represented in terms of the level spacing distribution and the two-level correlation function exhibit scale-invariant behavior. Moreover, they proved to be independent of the type of boundary conditions, provided that it belongs to a given topology. We have verified that for sufficiently large sizes one does not need to average over different random systems with the same disorder $W$. This is due to ergodicity principle, which establishes an equivalence between the averaging over an ensemble of many systems and averaging over the energy of the single system.

We have also succeeded to determine several eigenstates of the unprecedented matrix sizes of million-by-million very close to the center of the spectrum by partly using the conventional version of the restarted Lanczos solver with the reorthogonalization [4]. This was performed on the single-processor workstation Alpha DEC3000-AXP within a period of few weeks ($\approx 2 \cdot 10^5$ CPU-seconds). We have investigated the probability distribution of the local amplitudes of the wavefunctions and the energy dependence of their mutual correlations at the transition. It is interesting to notice that the achieved scale of disordered Hamiltonians becomes now comparable with typical realistic sizes of quantum dots and quantum wires in many experimental situations, although the Anderson model of localization does not take electron-electron interactions into account. We have also applied similar optimization techniques for diagonalizing the Hubbard model, where in addition to the disorder a few spinless fermions can interact if they are on the nearest-neighboring lattice cites (short-range coupling). The sparseness of the matrix of the Hubbard model is, however, less advantageous compared to the non-interacting models.

This work was supported by the SFB-Project 508 "Quanten Materialien". I.Kh.Zh. thanks the DFG for financial support during his stay at the University of Hamburg.

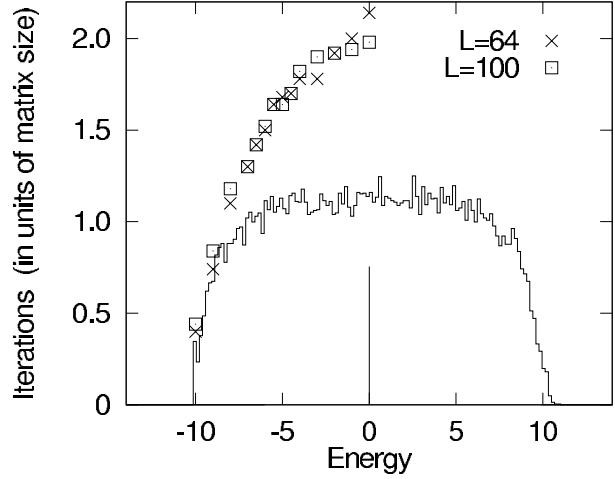

Figure 1: Energy dependence of the ratio, $M/N$, of $T_M$- and $A$-matrices for eigenpair to have accuracy $\delta = 10^{-9}$ at the disorder $W = 16.4$ for different lattice size $L$, $N = L^3$. Spectral density of states (rescaled), shown by full line, defines the band. Mobility edge is at the band center.